\documentclass[a4paper, 11pt]{article}  
\usepackage{geometry, graphicx, blindtext, authblk, subcaption}     
\usepackage{amssymb, amsmath}
\usepackage{algorithm} 
\usepackage{algpseudocode}
\usepackage[table]{xcolor}
\usepackage[colorlinks=true, linkcolor=blue, citecolor = blue]{hyperref}
\usepackage{booktabs, color}
\usepackage[toc,title]{appendix}
\usepackage{tikz}
\usepackage{cleveref} 
\usetikzlibrary{positioning}
\newcommand{\bx}{\mathbf{x}}

\newcommand{\Cov}{\mathbb{C}\text{ov}}
\newcommand{\Var}{\mathbb{V}}
\newcommand{\Exp}{\mathbb{E}}
\title{An ensemble-based approach for multi-fidelity emulation and adaptive sampling}
\author{Hossein Mohammadi\thanks{Corresponding Author: h.mohammadi@exeter.ac.uk} }
\affil{Faculty of Environment, Science and Economy, University of Exeter, UK}
\date{}							
\begin{document}
\maketitle
\begin{abstract}
High-resolution simulation models are essential for representing complex physical systems, yet their substantial computational cost severely limits the number of feasible high-fidelity (HF) evaluations. This problem is often addressed through multi-fidelity frameworks, which employ hierarchies of simulators with varying levels of fidelity and evaluation cost. A key difficulty in this setting is integrating information from such heterogeneous sources to accurately approximate HF simulators. This paper proposes a novel multi-fidelity emulation methodology based on ensemble learning. The base learners of the ensemble are hierarchical kriging emulators that systematically incorporate information from lower‑fidelity models into HF predictions. Aggregation of these base learners via Bayesian model averaging yields the multi-fidelity emulator with principled uncertainty quantification. The between-model variance component of this uncertainty is then employed as the acquisition criterion in an adaptive design strategy to enrich the training set with informative samples. The predictive performance of the approach is assessed on a collection of well-established benchmark problems. Results show that our multi-fidelity emulator outperforms single-model alternatives in terms of accuracy and robustness. Furthermore, the adaptive design strategy effectively identifies informative samples and improves emulator performance under limited computational budgets.
\end{abstract}
{\bf Keywords:} Ensemble learning; Gaussian processes; Hierarchical kriging; Multi-fidelity emulation
\section{Introduction}
\label{sec:introduction} 
The application of computer-based simulation models is ubiquitous across many scientific domains, ranging from climate science \cite{chau2021} to finance \cite{gonzalvez2019}. These simulators reproduce complex physical systems through intricate mathematical equations (e.g., partial or ordinary differential equations) that would otherwise be impossible to study. However, no computational tool can fully capture every aspect of reality. The accuracy of a simulator depends on the level of detail it incorporates. Models with greater complexity and finer resolution are considered high-fidelity (HF), while those with simplified representations are classified as low-fidelity (LF).

Although HF simulations are highly accurate, they consume substantial resources in terms of development time, computational overhead, and energy. To manage this burden, hierarchies of models are often employed. Such a hierarchy consists of multiple simulators describing the same quantity of interest (QoI), operating at different fidelity levels. Within this framework, the majority of simulation runs are conducted using LF models, which are significantly faster, while HF runs are reserved for a small number of critical evaluations. Readers are referred to \cite{peherstorfer2018} for a comprehensive survey of multi-fidelity methods and their applications.

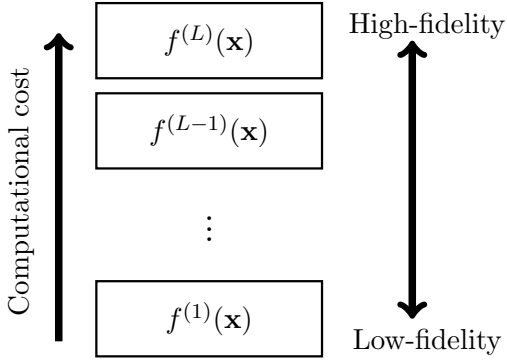
\begin{figure}[h]
	\centering
	\begin{tikzpicture}
		\draw[line width=0.3mm] (0.5, 0) rectangle (3.5, 1); 
		\node[] at (2,0.5)  {$f^{(1)}(\bx)$};
		\draw[line width=0.3mm] (0.5, 3.5) rectangle (3.5, 2.5); 
		\node[] at (2,3)  {$f^{(L-1)}(\bx)$};
		\node[] at (2,1.8)  {\Large\vdots};
		\draw[line width=0.3mm] (0.5, 3.7) rectangle (3.5, 4.7); 
		\node[] at (2,4.2)  {$f^{(L)}(\bx)$};
		\draw[->, line width=1mm]  (0, 0.2) -- (0, 4.3);
		\node [rotate=90] at (-0.5 ,2.2) {Computational cost};
		\draw[<->, line width=1mm]  (4.7, 0.5) -- (4.7, 4.2);
		\node [] at (4.9 ,4.4) {High-fidelity};
		\node [] at (4.9 ,0.2) {Low-fidelity};
	\end{tikzpicture}
	\caption{Hierarchy of computer codes with $L$ levels of fidelity. The HF simulator, denoted by $f^{(L)}(\bx)$, yields the most accurate representation of the QoI at a high computational cost. The lower-fidelity models, $f^{(1)}(\bx), \ldots, f^{(L-1)}(\bx)$, provide cheaper but less accurate approximations of the same QoI.} 
	\label{Fig:hierarchy_of_models}
\end{figure}

A hierarchy of models is illustrated schematically in \Cref{Fig:hierarchy_of_models}. The HF simulator is represented by the function $f^{(L)}(\bx)$, where $f^{(L)} : \mathcal X \mapsto \mathbb R$ and $\mathcal X\subseteq \mathbb R^d$. This model provides the most accurate estimate of the QoI but is computationally expensive. The lower-fidelity models are denoted by $f^{(1)}(\bx), \ldots, f^{(L-1)}(\bx)$, where $f^{(l)}  : \mathcal X \mapsto \mathbb R, l= 1, \ldots, L-1$. They are simplified versions of $f^{(L)}(\bx)$ and are therefore more computationally efficient. They also contain useful information about the underlying behaviour and dominant trends of the HF response.

Real-time prediction of HF simulators is vital in applications such as climate change and nuclear safety, where timely, informed decisions are required to protect lives, critical infrastructure, and the environment. Statistical emulators (or surrogate models) \cite{khuri2010, gramacy2020} are data‑driven approximations of expensive simulations and can be evaluated orders of magnitude faster. An emulator approximates the behaviour of a simulator by learning the input–output relationship from a limited number of simulation runs. Gaussian process (GP) emulators \cite{GPML} are widely used in this context. Their popularity stems from their favourable properties, including computational tractability and built‑in uncertainty quantification. However, constructing a reliable GP for HF models is challenging due to the extreme sparsity of available training data.

To tackle this issue, multi-fidelity emulation is commonly employed, which leverages supplementary information from lower-fidelity sources to construct an accurate emulator for the HF model. Various multi-fidelity techniques have been developed, a review of which can be found in \cite{brevault2020}. However, most existing approaches rely on the autoregressive framework proposed by Kennedy and O’Hagan (KOH) \cite{kennedy2000}, which adapts cokriging \cite{cressie2015} for computer experiments. Despite its widespread adoption, the KOH method suffers from several limitations, notably:
\begin{itemize}
	\item First, it typically requires \emph{nested experimental designs}, where HF points are a subset of the LF data. This requirement restricts its applicability in practice, as many real-world datasets consist of scattered, non-nested observations.

	\item Second, cokriging involves the inversion of a large joint covariance matrix, which is computationally demanding and can give rise to numerical robustness issues, such as ill-conditioning. Although this issue has been mitigated by the recursive formulation of the KOH method \cite{legratiet2013, legratiet2014}, the theoretical exactness of this approach still relies on the availability of nested designs. 
	
	\item Third, as discussed in \Cref{sec:HK_model}, the correlation between simulators in the KOH framework is governed by a single scalar coefficient, which can severely restrict model flexibility. To address this problem, recent work has explored non-linear correlation structures \cite{perdikaris2017, cutajar2019, ko2022, heo2025}, often through deep GPs \cite{damianou2013} that treat each fidelity level as a hidden layer. However, training such models typically needs advanced inference techniques, such as variational inference, which are computationally intensive and difficult to tune in data-sparse regimes.
\end{itemize}
 
Despite these advances, there remains a need for multi-fidelity emulators that combine flexibility, computational efficiency, and robust uncertainty quantification without restrictive design assumptions. To address these limitations, this paper proposes an ensemble-based multi-fidelity emulation framework. Ensemble methods offer a more flexible modelling strategy than single-model approaches by combining predictions from multiple surrogate models, referred to as base learners. Although ensemble learning has demonstrated strong performance in broader machine learning contexts \cite{acar2009, polikar2012, mienye2022}, its potential remains largely unexplored in multi-fidelity modelling. 

In this work, the base learners of the ensemble are hierarchical kriging (HK) emulators with distinct covariance functions \cite{han2012} (see \Cref{sec:HK_model}). HK is a GP–based model that enables information transfer across fidelity levels by treating the lower-fidelity response as a deterministic trend for the emulator at the next fidelity level. This mechanism allows efficient exploitation of information from lower-fidelity simulations. By employing HK as base learners, the proposed approach avoids limitations of conventional autoregressive methods, such as the requirement for nested experimental designs and the computational burden associated with inverting large joint covariance matrices. As a result, the framework is well suited to real-world applications where low- and high-fidelity observations may not overlap. 

The multi-fidelity emulator is obtained by aggregating these base learners via Bayesian model averaging (BMA) \cite{raftery1997, raftery2005}. BMA accounts for model uncertainty by combining the predictive distributions of the individual base learners according to their posterior probabilities, yielding a unified surrogate model with principled uncertainty quantification. This uncertainty is then used to guide adaptive sampling, also known as active learning \cite{liu2018, fuhg2021}. Adaptive strategies play a key role in the emulation of computationally expensive codes, since initial surrogate models may be unreliable due to limited training data. Accordingly, such strategies are employed to iteratively enrich the training set with informative samples. The main advantage of this sequential process is that it can be terminated once a desired level of predictive accuracy has been achieved. This flexibility is not available with traditional one-shot designs \cite{mohammadi2022}, where the sample size is fixed prior to model fitting. 

Adaptive strategies are guided by an acquisition criterion that directs sampling toward regions of the input space where additional evaluations are expected to provide the greatest information about the underlying function. In this work, the proposed acquisition criterion is based on the between-model variance derived from the BMA aggregation. Compared with the single-fidelity case, adaptive design in the multi-fidelity setting introduces an additional decision step: once a new input location has been selected, the corresponding fidelity level must be determined. This decision is made by maximising the ratio of expected information gain to computational cost, where the gain is quantified via variance reduction within the BMA framework.

The remainder of the paper is organised as follows. \Cref{sec:background} introduces the foundations of GPs, HK, and BMA. \Cref{sec:proposed_method} and \Cref{sec:adaptive_design} detail the proposed multi-fidelity emulator and adaptive sampling, respectively. In \Cref{sec:experimental_res}, the performance of our method is evaluated on a suite of test functions. Finally, \Cref{sec:conclusion} provides a summary of our findings and some concluding thoughts on future work.
\section{Technical background}
\label{sec:background}
This section details the theoretical foundations of the proposed multi-fidelity framework. We first review GP and HK emulators, followed by the BMA approach used to aggregate the ensemble members.
\subsection{Gaussian processes} 
\label{sec:GP_model}
A GP, denoted by $Y(\bx)$, is a stochastic process indexed by $\bx \in \mathcal{X} \subseteq \mathbb{R}^d$ such that any finite collection of its random variables follows a joint multivariate normal distribution \cite{GPML}. The process is fully specified by its mean (trend) function $\mu : \mathcal{X} \to \mathbb{R}$ and a symmetric positive-definite covariance function/kernel $k : \mathcal{X} \times \mathcal{X} \to \mathbb{R}$, defined as:
\begin{equation}
	k(\bx, \bx'; \boldsymbol{\theta}) = \Cov\left(Y(\bx), Y(\bx')\right),
\end{equation}
where $\boldsymbol{\theta}$ represents the vector of kernel hyperparameters. In the absence of prior information regarding the global trend, a constant mean $\mu(\bx) = \mu_0$ is typically assumed. Under this assumption, the process is expressed as:
\begin{equation}
	Y(\bx) = \mu_0 + Z(\bx),
	\label{GP_prior}
\end{equation}
where $Z(\bx)$ is a zero-mean GP. 

Given a training set $\mathcal{D} = \{ (\bx_i, f(\bx_i))\}_{i = 1}^n$, the posterior process $Y(\bx) \mid \mathcal{D}$ remains a GP. By conditioning the prior on the observed vector $\mathbf{f} = (f(\bx_1), \dots, f(\bx_n))^\top$, the posterior predictive mean and variance are given by:
\begin{align} 
	\label{E:kriging_mean}
	m(\bx) &= \mu_0 + \mathbf{k}(\bx)^\top \mathbf{K}^{-1}(\mathbf{f} - \mu_0 \mathbf{1}), \\
	\sigma^2(\bx) &= k(\bx, \bx) - \mathbf{k}(\bx)^\top \mathbf{K}^{-1}\mathbf{k}(\bx),
	\label{E:kriging_var}
\end{align}
where $\mathbf{k}(\bx) = \left(k(\bx, \bx_1), \ldots, k(\bx, \bx_n)\right)^\top$, $\mathbf{K}$ is the vector of cross-covariances between the test point and training points, $\mathbf{K}$ is the $n \times n$ covariance matrix with entries $K_{ij} = k(\bx_i, \bx_j)$, and $\mathbf{1}$ is a vector of ones.

The hyperparameters $\boldsymbol{\theta}$ and the trend $\mu_0$ are generally unknown and are estimated by maximising the log marginal likelihood:
\begin{equation}
	\mathcal{L}(\boldsymbol{\theta}) = -\frac{1}{2}(\mathbf{f} - \mu_0 \mathbf{1})^\top \mathbf{K}^{-1}(\mathbf{f} - \mu_0 \mathbf{1}) - \frac{1}{2} \log \det \mathbf{K} - \frac{n}{2}\log 2\pi.
	\label{log-lik}
\end{equation}
The resulting maximum likelihood estimates $\hat{\boldsymbol{\theta}}$ and $\hat{\mu}_0$, are substituted into \eqref{E:kriging_mean}--\eqref{E:kriging_var} to perform prediction.
\subsection{Hierarchical kriging}
\label{sec:HK_model}
HK \cite{han2012} provides a rigorous yet simple framework for incorporating information from lower-fidelity codes. The idea is that the predictive mean of a lower-fidelity model is treated as the deterministic trend function for the subsequent higher-fidelity emulator. Consider the hierarchy of models illustrated in \Cref{Fig:hierarchy_of_models}. Let $m^{(l-1)}(\bx)$ denote the predictive mean of the GP fitted to the $(l-1)$-th fidelity model. The HK emulator for the $l$-th fidelity model, $f^{(l)}(\bx)$, is defined as
\begin{equation}
	Y^{(l)}(\bx) = \beta^{(l)} \, m^{(l-1)}(\bx) + Z^{(l)}(\bx),
\end{equation}  
where $\beta^{(l)}$ is a scaling factor estimated via maximum likelihood, and $Z^{(l)}(\bx)$ is a zero-mean GP with covariance function $k^{(l)}(\bx, \bx')$. The factor $\beta^{(l)}$ reflects the degree of correlation between $f^{(l-1)}$ and $f^{(l)}$ \cite{han2012}.

The HK prediction at level $l$ is obtained by conditioning $Y^{(l)}(\bx)$ on the training set $\mathcal{D}^{(l)}$, resulting in the posterior mean $m^{(l)}(\bx)$ and variance $\sigma^{2(l)}(\bx)$. A distinctive feature of the HK model is its ability to explicitly account for the uncertainty propagation from lower-fidelity levels. Specifically, the predictive variance $\sigma^{2(l)}(\bx)$ comprises the intrinsic uncertainty of $Z^{(l)}$ and the propagated uncertainty from the previous level. The contribution of the $(l-1)$-th emulator to the total uncertainty at level $l$ is given by $(\beta^{(l)})^2 \sigma^{2(l-1)}(\bx)$ \cite{zhang2018_HK}. This recursive formulation implies that an observation at level $l-1$ reduces the posterior HK variance at level $l$ by this specific propagated term. This property is fundamental to the fidelity level selection logic described in \Cref{sec:adaptive_design}.

For the purpose of comparison, the KOH method is briefly introduced here to highlight its differences from HK. In KOH, the emulator for $f^{(l)}(\bx)$ is defined by the following autoregressive relation:
\begin{equation}
	Y^{(l)}(\bx) = \rho Y^{(l-1)}(\bx) + \delta(\bx),
\end{equation}  
where $\rho$ is a scalar and $\delta(\bx)$ is a discrepancy GP that captures the difference between fidelity levels. This formulation requires estimating the cross-covariance between $Y^{(l-1)}(\bx)$ and $\delta(\bx)$, which leads to a significantly larger covariance matrix and potential issues regarding numerical stability. As previously discussed, HK allows for a more straightforward implementation and reduced computational overhead by circumventing such joint modelling.
\subsection{Bayesian model averaging}
\label{sec:BMA}
BMA \cite{raftery1997,raftery2005} is a principled approach for aggregating predictions from multiple surrogate models. The resulting predictive distribution is obtained as a weighted combination of the individual model predictions, with weights given by posterior model probabilities. Let $M_1, \ldots, M_S$ denote an ensemble of surrogate models for a QoI $\Delta$. The predictive distribution under BMA is given by:
\begin{equation}
	p\left(\Delta \mid \mathcal{D}\right) = \sum_{s = 1}^{S}  p\left(\Delta \mid \mathcal{D}, M_s\right) \, p\left(M_s \mid \mathcal{D}\right),
\end{equation}
where $p\left(M_s\mid \mathcal{D}\right) = \omega_s$ is the posterior probability of model $M_s$. By Bayes’ theorem, these weights are defined as:
\begin{equation}
	\omega_s = \frac{p\left(\mathcal{D} \mid M_s \right) p\left(M_s\right)}
	{\sum_{t = 1}^S p\left(\mathcal{D} \mid M_t \right) p\left(M_t\right)},
	\label{BMA_weight}
\end{equation}
where $p(\mathcal{D} \mid M_s)$ is the marginal likelihood and $p(M_s)$ is the prior model probability. These weights are non-negative and satisfy the normalisation constraint $\sum_{s = 1}^S \omega_s  = 1$. 

Let $\mu_s$ and $\sigma_s^2$ denote the predictive mean and variance under model $M_s$, respectively. The BMA predictive mean and variance are given by \cite{raftery1997,duan2007}:
\begin{align}
	&\Exp\left[\Delta \mid \mathcal{D}\right] = \sum_{s = 1}^{S} \omega_s  \mu_s , \\
	&\Var\left(\Delta \mid \mathcal{D}\right) = \underbrace{\sum_{s = 1}^{S}  \omega_s \sigma^2_s}_{\text{within-model variance}}  +  \underbrace{\sum_{s = 1}^{S}  \omega_s \left(\mu_s -  \Exp\left[\Delta \mid \mathcal{D}\right] \right)^2}_{\text{between-model variance}} . 
	\label{BMA_Var}
\end{align}
The first term on the right-hand side of \Cref{BMA_Var} represents the within-model variance, which accounts for the average epistemic uncertainty of the individual base learners due to a lack of data. The second term, the between-model variance, quantifies the structural disagreement (or ambiguity) among the ensemble members. This component accounts for model misspecification by capturing the uncertainty inherent in the selection of the model architecture itself. This between‑model variance forms the basis of the adaptive design strategy introduced in \Cref{sec:adaptive_design}.

In this work, each $M_s$ is a GP emulator with a distinct covariance function. Assuming equal prior probabilities, $p(M_s) = 1/S$, the weights in \Cref{BMA_weight} are proportional to the marginal likelihoods. Using the plug-in hyperparameter estimates $\hat{\boldsymbol{\theta}}_s$ as defined in \eqref{log-lik}, the weights are calculated as:
\begin{equation}
	\omega_s = \frac{\exp \left(\mathcal{L}\left(\hat{\boldsymbol{\theta}}_s\right)\right)}
	{\sum_{t = 1}^S \exp \left(\mathcal{L}\left(\hat{\boldsymbol{\theta}}_t\right)\right)}, 
	\label{BMA_weight_L}
\end{equation}
where $\mathcal{L}\left(\hat{\boldsymbol{\theta}}_s\right)$ is the log marginal likelihood of the $s$-th GP. In this way, emulators that achieve a better fit to the observed data receive greater weight.
\section{Multi-fidelity emulator}
\label{sec:proposed_method}
This section introduces the proposed multi-fidelity emulator, which relies on ensemble learning. The base learners of the ensemble consist of HK models with distinct covariance functions, constructed for the HF simulator. They are then aggregated using BMA to obtained a unified predictor with associated uncertainty quantification. The underlying motivation is that ensembles of surrogate models often achieve higher predictive accuracy than individual surrogates \cite{acar2009, polikar2012, mienye2022}. This is particularly critical in sparse-data regimes where HF observations are limited. Also, the predictive variance offered by BMA provides a natural foundation for adaptive design to further improve the emulator.

Let $\mathcal{D}^{(l)}$ be the training set associated with $f^{(l)}$, the model at fidelity level $l$, for $l = 1, \ldots, L$. The procedure begins at the lowest level ($l = 1$), where an ensemble of $S$ GP emulators is constructed using $\mathcal{D}^{(1)}$ and $S$ distinct covariance kernels. The resulting predictive means, $\{m^{(1)}_s(\bx)\}_{s=1}^S$, are then employed as deterministic trend functions for the emulators at the subsequent fidelity level. For $l = 2, \ldots, L$, the $S$ HKs are defined as:
\begin{equation}
	Y^{(l)}_s(\bx) = \beta_s^{(l)} m^{(l-1)}_s(\bx) + Z^{(l)}_s(\bx), \quad s = 1, \ldots, S,
\end{equation}
where $\beta_s^{(l)}$ is a scaling coefficient and $Z^{(l)}_s(\bx)$ is a zero-mean GP. This recursive structure ensures that each base learner at level $l$ inherits the spatial information captured at level $l-1$. Note that since $m^{(l-1)}_s(\bx)$ is a deterministic continuous function defined over the entire input domain, the training points in $\mathcal{D}^{(l)}$ can be located at any arbitrary coordinates. This eliminates the need for nested experimental designs and allows the framework to fuse information from scattered datasets.

At the highest fidelity level $L$, the ensemble consists of $S$ base learners with posterior distributions:
\begin{equation}
	Y^{(L)}_s(\bx) \mid \mathcal{D}^{(L)} \sim \mathcal{N}\left(m^{(L)}_s(\bx), \sigma^{2(L)}_s(\bx)\right), \quad s = 1, \ldots, S.
\end{equation}
These $S$ base learners are aggregated through BMA to construct the multi-fidelity emulator $\text{MF}(\bx)$. The predictive mean $\bar{m}(\bx)$ and variance $\bar{\sigma}^2(\bx)$ of $\text{MF}(\bx)$ are given by:
\begin{align}
	\label{ensemble_mean}
	&\bar{m}(\bx) = \sum_{s= 1}^S \omega_s m^{(L)}_s(\bx), \\
	&\bar{\sigma}^2(\bx) = \underbrace{\sum_{s= 1}^S\omega_s \sigma^{2(L)}_s(\bx)}_{\sigma^2_{\mathrm{wm}}(\bx)} + \underbrace{\sum_{s= 1}^S \omega_s \left(m^{(L)}_s(\bx) - \bar{m}(\bx)\right)^2}_{\sigma^2_{\mathrm{bm}}(\bx)},
	\label{ensemble_var}
\end{align}
where $\omega_s$ are the BMA weights, and $\sigma^2_{\mathrm{wm}}(\bx)$ and $\sigma^2_{\mathrm{bm}}(\bx)$ represent the within- and between-model variances, respectively. As explained in \Cref{sec:BMA}, the weights $\omega_s$ are proportional to the marginal likelihood of each base learner, normalised to sum to unity, under the assumption of equal prior probabilities. \Cref{alg:BMA_HK} outlines the implementation steps of the proposed multi-fidelity emulation. 

\begin{algorithm}
	\caption{Multi-fidelity emulation algorithm}
	\label{alg:BMA_HK}
	\begin{algorithmic}[1]
		\State \textbf{Input:} Training data sets $\{\mathcal{D}^{(l)}\}_{l=1}^L$ and $S$ distinct covariance kernels.
		\State \textbf{Output:} Multi-fidelity predictive mean $\bar{m}(\bx)$ and variance $\bar{\sigma}^2(\bx)$.
		\For{$s = 1$ to $S$}
		\State Construct GP emulator $Y^{(1)}_s$ using $\mathcal{D}^{(1)}$ and the $s$-th kernel;
		\State Extract the posterior mean function $m^{(1)}_s(\bx)$;
		\EndFor
		\For{$l = 2$ to $L$}
		\For{$s = 1$ to $S$}
		\State Define $Y^{(l)}_s(\bx) = \beta_s^{(l)} m^{(l-1)}_s(\bx) + Z^{(l)}_s(\bx)$;
		\State Use $\mathcal{D}^{(l)}$ to estimate $\beta_s^{(l)}$ and hyperparameters of $Z^{(l)}_s$;
		\State Store posterior mean $m^{(l)}_s(\bx)$ and variance $\sigma^{2(l)}_s(\bx)$;
		\EndFor
		\EndFor
		\State Aggregate $Y^{(L)}_s(\bx) \mid \mathcal{D}^{(L)}, s = 1, \ldots, S,$ through BMA;
		\State Compute $\bar{m}(\bx)$ and $\bar{\sigma}^2(\bx)$ using Equations (\ref{ensemble_mean}) and (\ref{ensemble_var}), respectively;
		
		\State \Return $\bar{m}(\bx), \bar{\sigma}^2(\bx)$
	\end{algorithmic}
\end{algorithm}

\begin{figure}[htbp]
	\centering
	\includegraphics[width=0.48\textwidth]{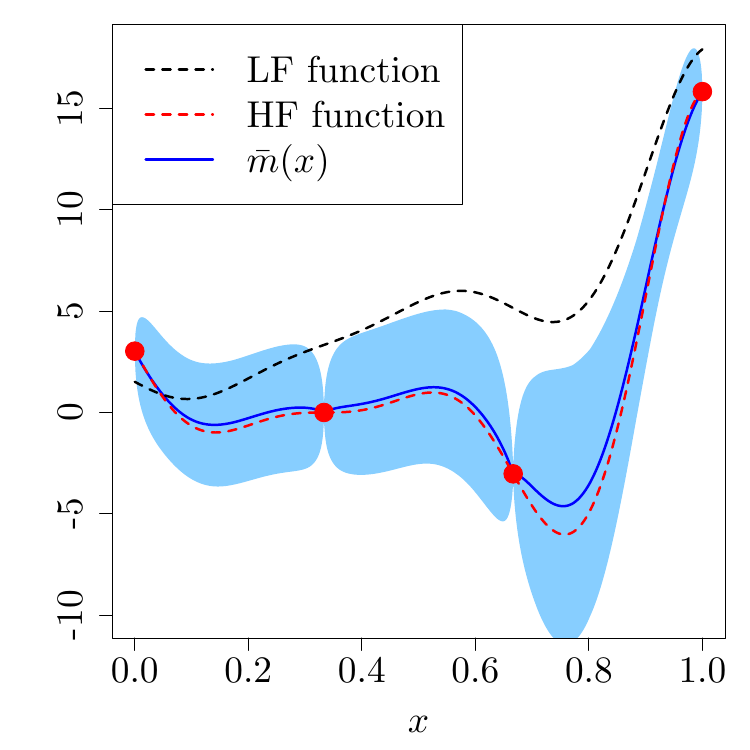}
	\includegraphics[width=0.48\textwidth]{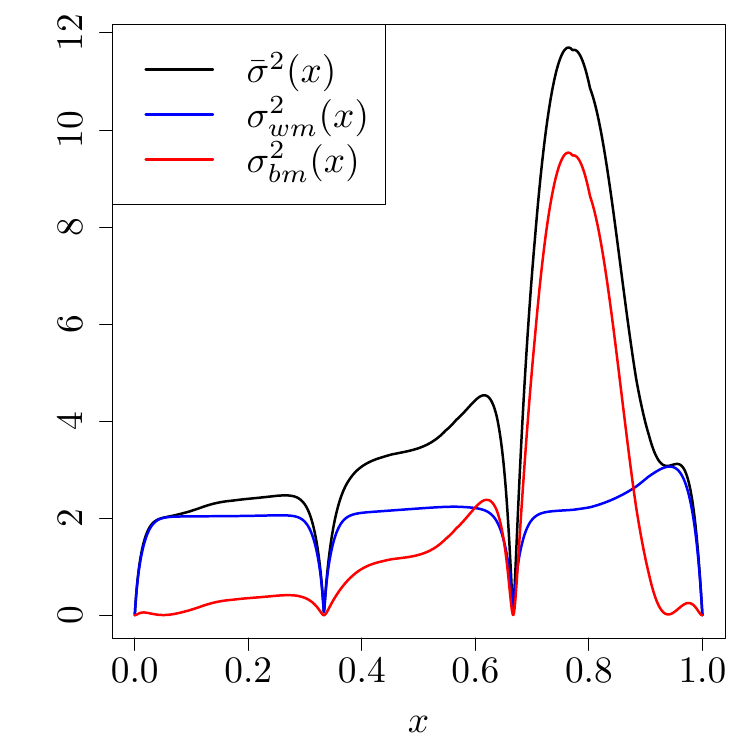}
	\caption{Left: Multi-fidelity predictive mean $\bar{m}(\bx)$ (solid blue) together with the associated uncertainty bounds $\bar{m}(\bx) \pm 2\bar{\sigma}(\bx)$ (shaded area), shown relative to the LF (black dashed) and HF (red dashed) functions. For visual clarity, the LF points are omitted. Right: Multi-fidelity predictive variance $\bar{\sigma}^2(\bx)$ (black) which is the sum of the within-model variance $\sigma^2_{\mathrm{wm}}(\bx)$ (blue) and the between-model variance $\sigma^2_{\mathrm{bm}}(\bx)$ (red).}
	\label{forrester_fun}
\end{figure}

It is worth noting that $\text{MF}(\bx)$ interpolates the HF observations; its mean $\bar{m}(\bx)$ passes through these points and its variance $\bar{\sigma}^2(\bx)$ vanishes. This follows from the fact that each base learner interpolates the HF data and the BMA weights sum to unity. \Cref{forrester_fun} illustrates the multi-fidelity emulation of a function defined in the 1D space. The left panel shows the LF (black dashed) and HF (red dashed) functions, the multi-fidelity predictive mean (blue) and the associated uncertainty bounds (shaded area), $\bar{m}(\bx) \pm 2\bar{\sigma}(\bx)$. The right panel displays the multi-fidelity predictive variance $\bar{\sigma}^2(\bx)$ (black) which consists of the within-model variance $\sigma^2_{\mathrm{wm}}(\bx)$ (blue) and the between-model variance $\sigma^2_{\mathrm{bm}}(\bx)$ (red). These components are used in the adaptive sampling strategy presented in the next section to determine both the location and the fidelity of future evaluations.

\section{Multi-fidelity adaptive sampling}
\label{sec:adaptive_design}
The initial accuracy of multi-fidelity emulators is often limited by the sparsity of available HF data. To address this, adaptive design strategies are employed to sequentially enrich the training set with informative samples. This iterative procedure is governed by an acquisition criterion, which identifies the optimal locations for subsequent sampling. Compared with the single-fidelity case (see, e.g., \cite{mohammadi2022, mohammadi2025}), adaptive design for multi-fidelity models require an additional decision-making step: determining the most appropriate fidelity level $l \in \{1, \ldots, L\}$ at which to evaluate each newly chosen input location.

In this work, the between-model variance $\sigma^2_{\mathrm{bm}}(\bx)$ defined in \Cref{ensemble_var} serves as the acquisition criterion. In the active learning literature, sampling strategies based on $\sigma^2_{\mathrm{bm}}(\bx)$ are commonly referred to as \emph{query-by-committee} approaches \cite{burbidge2007, liu2018}. At each iteration, the next candidate location $\bx^\ast$ is determined by maximising this variance:
\begin{equation}
	\bx^\ast = \underset{\bx \in \mathcal{X}_{\mathrm{cand}}}{\arg\max} ~ \sigma^2_{\mathrm{bm}}(\bx).
	\label{eq:ensemble_var_max}
\end{equation}
This optimisation is carried out by performing a discrete search over a set of candidate points $\mathcal{X}_{\mathrm{cand}}$. Adopting a discrete search, rather than continuous optimisation, reduces the risk of becoming trapped in local optima and avoids the computational overhead associated with evaluating gradients of the ensemble variance. To ensure representative coverage of the input space, the set $\mathcal{X}_{\mathrm{cand}}$ is generated using a space-filling design, such as Latin hypercube sampling (LHS) \cite{pronzato2012}.

The effectiveness of $\sigma^2_{\mathrm{bm}}(\bx)$ as an acquisition criterion can be explained via the \emph{ambiguity decomposition} \cite{krogh1995, mendes2012}:
\begin{equation}
	\left(\bar{m}(\bx) - f^{(L)}(\bx)\right)^2 = \sum_{s=1}^{S} \omega_s \left(m_s^{(L)}(\bx) - f^{(L)}(\bx)\right)^2 - \sigma^2_{\mathrm{bm}}(\bx).
	\label{eq:ambiguity_decom}
\end{equation}
This decomposition relates the squared error of the ensemble mean prediction (left-hand side) to two components. The first term on the right-hand side is the weighted average of the squared errors of the individual base learners, while the second term is the between-model variance that quantifies disagreement (or ambiguity) among ensemble members. By maximising $\sigma^2_{\mathrm{bm}}(\bx)$, the proposed acquisition strategy selects the input location $\bx^\ast$ at which the ensemble members reach the greatest disagreement. Since $\sigma^2_{\mathrm{bm}}(\bx) \ge 0$, a larger ambiguity term implies a smaller ensemble prediction error for a fixed average errors of the base learners. Therefore, sampling at $\bx^\ast$ can effectively reduce the overall prediction error of the ensemble. 

It is also observed that using $\sigma^2_{\mathrm{bm}}(\bx)$ as the acquisition criterion steers sampling towards ``interesting" regions, such as areas where the model output exhibits strong non-linearity or abrupt changes. This behaviour can not be achieved when using the ensemble predictive variance $\bar{\sigma}^2(\bx)$. The reason is that $\bar{\sigma}^2(\bx)$ yields a global measure of uncertainty, dominated by $\sigma^2_{\mathrm{wm}}(\bx)$, which tends to peak in regions far from existing training data. As a result, strategies based on $\bar{\sigma}^2(\bx)$ promote global exploration and often lead to space‑filling designs, with a substantial number of samples on the boundaries of the input space.

\begin{figure}[htpb] 
	\centering
	\includegraphics[width=0.48\textwidth]{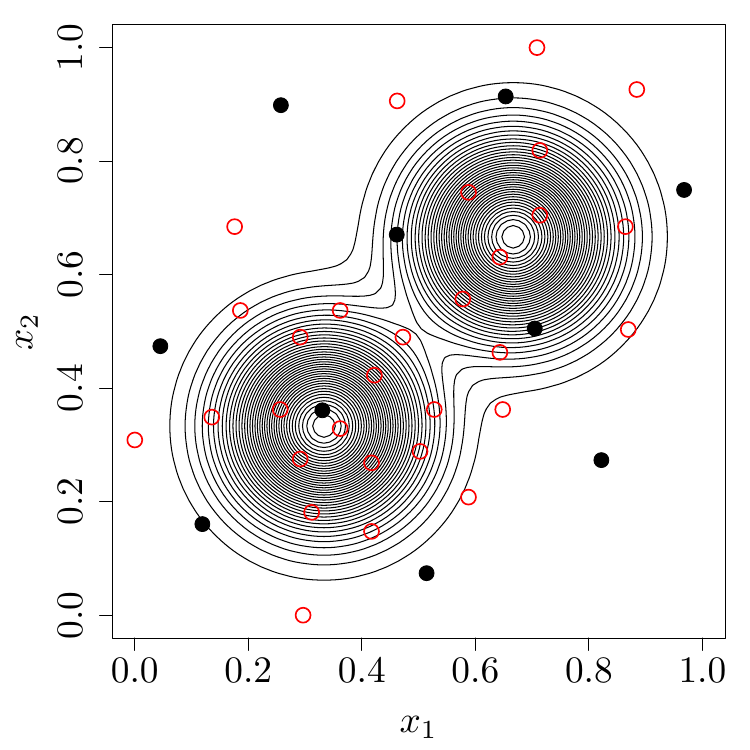}
	\includegraphics[width=0.48\textwidth]{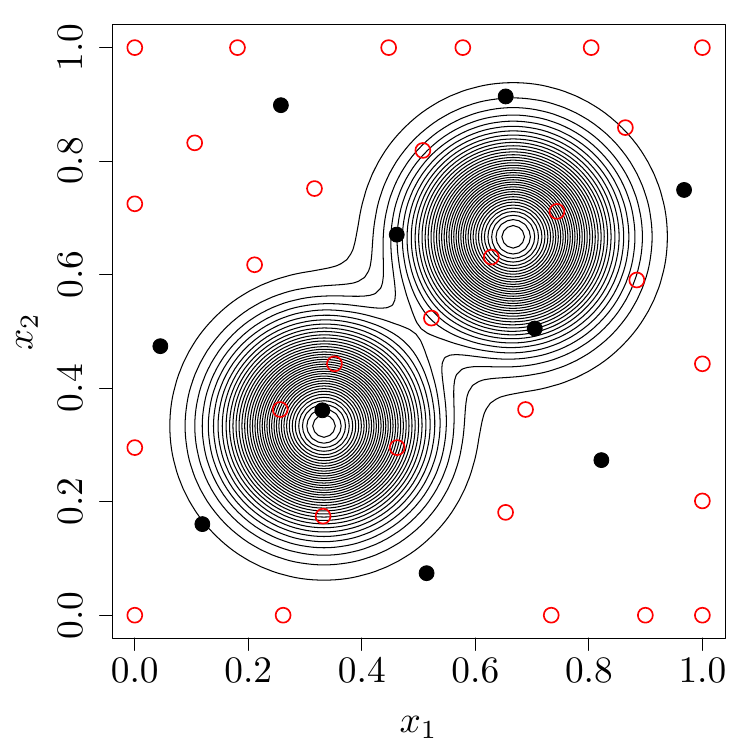}  
	\includegraphics[width=0.48\textwidth]{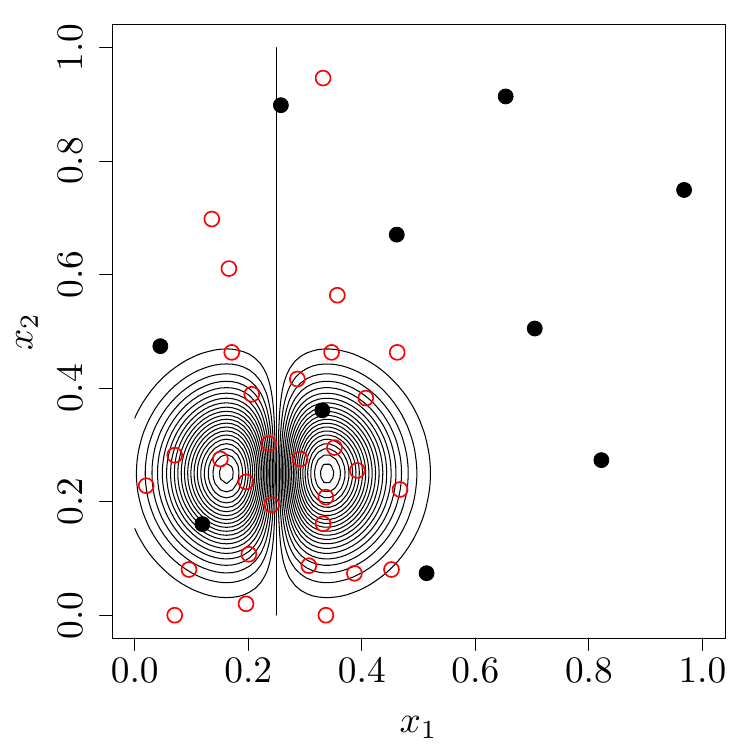}
	\includegraphics[width=0.48\textwidth]{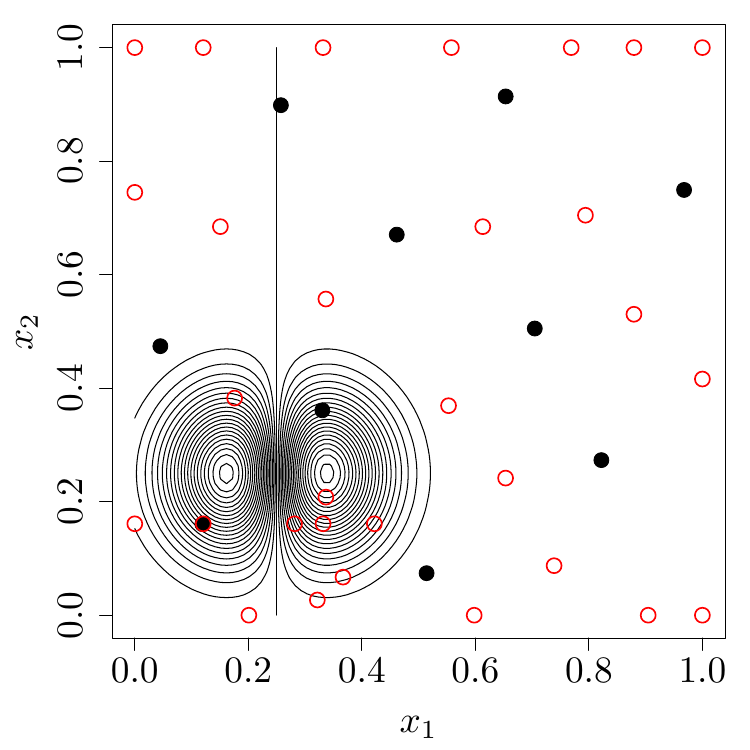}  
\caption{Adaptive sampling driven by the between-model variance $\sigma^2_{\mathrm{bm}}(\bx)$ (left) and the multi-fidelity predictive variance $\bar{\sigma}^2(\bx)$ (right). Black dots denote the initial design points and red circles are samples selected during the adaptive process. The criterion $\bar{\sigma}^2(\bx)$ promotes global exploration, leading to space-filling behaviour and concentration of samples near the boundaries of the input space. In contrast, $\sigma^2_{\mathrm{bm}}(\bx)$ directs sampling towards regions of high non-linearity or abrupt changes, where disagreement among ensemble members is greatest.}
	\label{bm_var_perform}
\end{figure}

By employing $\sigma^2_{\mathrm{bm}}(\bx)$ instead, the adaptive sampling targets locations where the disagreement among the base learners is largest. This allows the computational budget to be spent on resolving structural model misspecification rather than simply filling gaps in the design space. To demonstrate this behaviour, \Cref{bm_var_perform} compares the sampling patterns obtained by $\sigma^2_{\mathrm{wm}}(\bx)$ (left) and $\bar{\sigma}^2(\bx)$ (right) using two single-fidelity examples. Initial design points and adaptively selected samples are shown by black dots and red circles, respectively.

Once the candidate location $\bx^\ast$ is selected, the optimal fidelity level $l^\ast$ must be identified. This is typically determined by evaluating the expected information gain per unit of computational cost \cite{remilam2015}. To this end, the information gain metric, $\tilde{\sigma}^2(\bx^\ast, l)$ is defined as the expected reduction in predictive uncertainty at the highest fidelity level $L$ that would result from an evaluation at $\bx^\ast$ and fidelity level $l$. This quantity is given by:
\begin{equation}
	\tilde{\sigma}^2(\bx^\ast, l) = 
	\begin{cases}
		\bar{\sigma}^2(\bx^\ast), & l = L, \\
		\sum_{s=1}^{S} \omega_s B_s^{(l)} \sigma_s^{2(l)}(\bx^\ast), & l < L,
	\end{cases}
\end{equation}
where $B_s^{(l)} = \prod_{t=l+1}^{L} (\beta_s^{(t)})^2$ \cite{zhang2024}. At the highest level ($l=L$), the information gain is simply equal to the predictive variance of the multi-fidelity emulator, $\bar{\sigma}^2(\bx^\ast)$. For lower-fidelity levels ($l < L$), the proposed metric captures the reduction in uncertainty propagated through the hierarchy, accounting for the contribution of the lower-fidelity trend components to the highest-fidelity prediction.

Finally, the optimal fidelity level $l^\ast$ is obtained by solving the following optimisation problem:
\begin{equation}
	l^\ast = \underset{l \in \{1, \ldots, L\}}{\arg\max} \, \frac{\tilde{\sigma}^2(\bx^\ast, l)}{C^{(l)}},
\end{equation}
where $C^{(l)}$ denotes the computational cost associated with evaluating the simulator at fidelity level $l$.
\section{Experimental results}
\label{sec:experimental_res}
This section describes the experimental setups used to assess the predictive performance of the multi-fidelity emulator. The ensemble consists of $S=4$ base learners, each corresponding to a HK model with a distinct stationary covariance function. These kernels are selected to represent different levels of smoothness of the underlying process: the squared exponential (infinitely differentiable), Mat\'{e}rn 5/2 (twice differentiable), Mat\'{e}rn 3/2 (once differentiable), and exponential (non-differentiable) kernels. This set covers the most widely used covariance structures in the GP literature. Furthermore, they are readily available in standard GP software environments, including the \textsf{R} package \texttt{DiceKriging} \cite{roustant2012}.

Predictive accuracy is quantified via the root mean squared error (RMSE) and benchmarked against the recursive cokriging \cite{legratiet2014} and recursive non-additive (RNA) approaches \cite{heo2025}. They are implemented in the \texttt{MuFiCokriging} and \texttt{RNAmf} packages, respectively, with both employing a Mat\'{e}rn 5/2 covariance function. All methods are evaluated on the following well-recognised benchmark functions (see \cite{bingham2024} for more details):
\begin{itemize}
	\item Currin function (2D);
	\item Park function 1 (4D);
	\item Park function 2 (4D);
	\item Hartmann function (6D).
\end{itemize}

For the purpose of clarity and to facilitate analysis of the proposed methodology, a two-level hierarchy ($L = 2$) is considered for the test cases. This setup represents a common industrial scenario where a computationally expensive HF code is augmented by a cheaper LF approximation. The cost of evaluating the HF simulator is assumed to be seven times higher than that of the LF model. Analytical expressions for these test functions are provided in Appendix \ref{app:test_functions}.

\begin{figure}[htbp]
	\centering
	\begin{subfigure}[b]{0.48\textwidth}
		\centering
		\includegraphics[width=\textwidth]{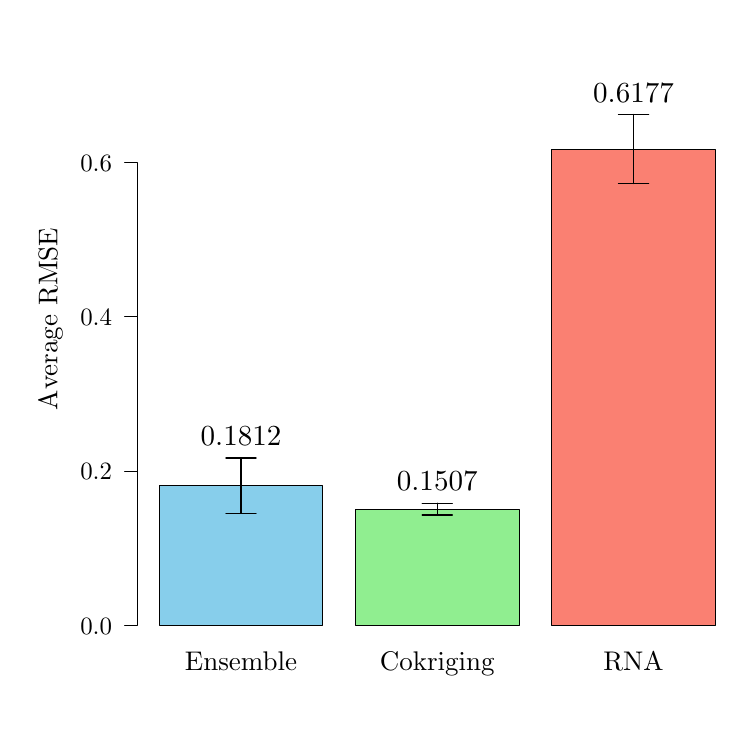}
		\caption{High sparsity, Nested}
		\label{fig:currin_ve_nest}
	\end{subfigure}
	\hfill
	\begin{subfigure}[b]{0.48\textwidth}
		\centering
		\includegraphics[width=\textwidth]{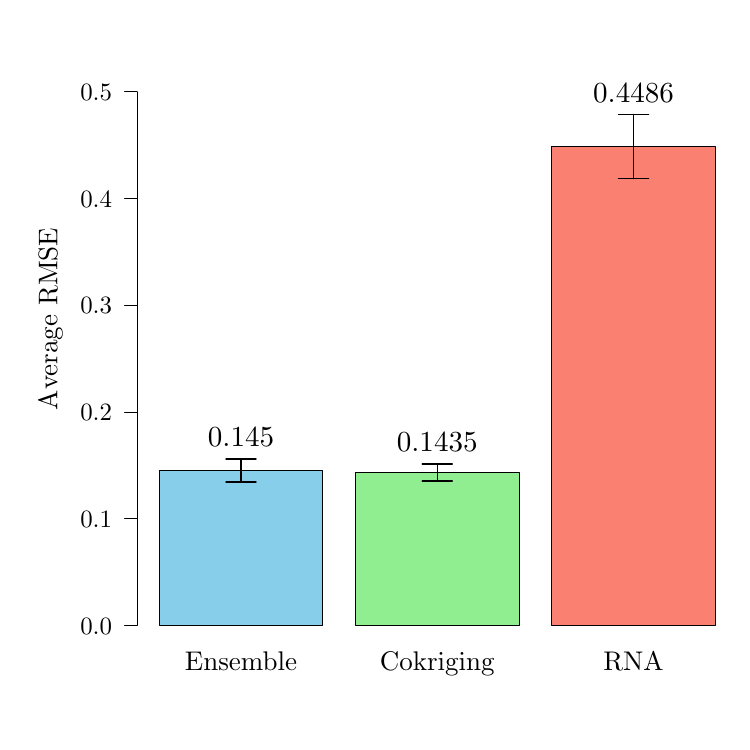}
		\caption{High sparsity, Non-nested}
		\label{fig:currin_ve_non}
	\end{subfigure}
	
	\vspace{1em}

	\begin{subfigure}[b]{0.48\textwidth}
		\centering
		\includegraphics[width=\textwidth]{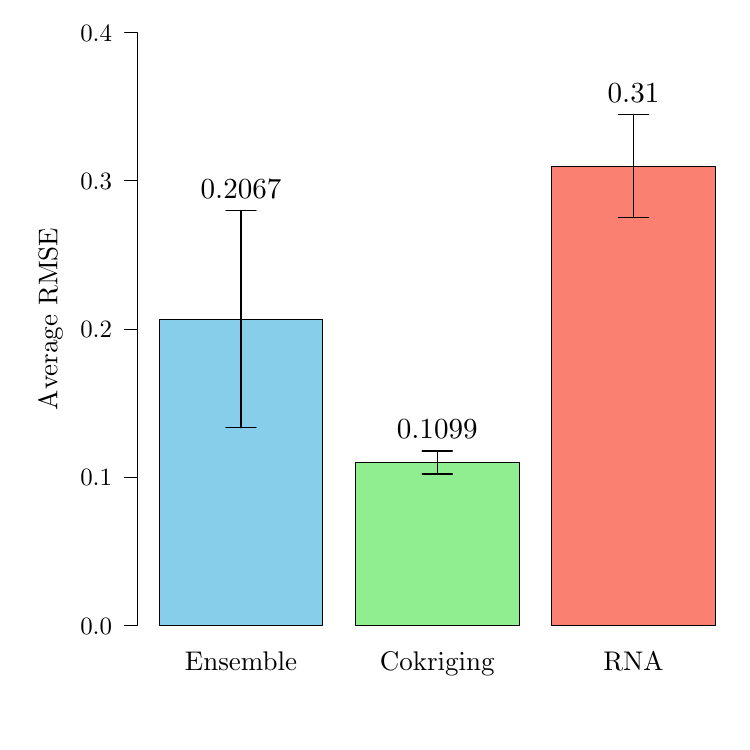}
		\caption{Low sparsity, Nested}
		\label{fig:currin_e_nest}
	\end{subfigure}
	\hfill
	\begin{subfigure}[b]{0.48\textwidth}
		\centering
		\includegraphics[width=\textwidth]{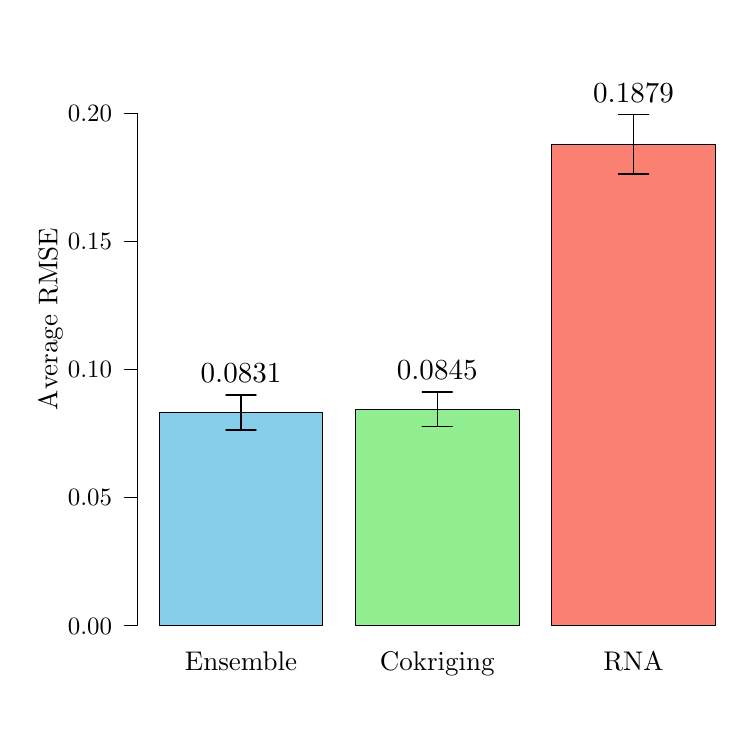}
		\caption{Low sparsity, Non-nested}
		\label{fig:currin_e_non}
	\end{subfigure}
	
	\caption{Average RMSE for the 2D Currin function evaluated over 30 different initial LHS designs under low and high sparsity regimes and nested and non-nested sampling settings. ``Ensemble", ``Cokriging", and ``RNA" are referred to the proposed framework, recursive cokriging, and recursive non-additive approaches, respectively.}
	\label{fig:currin_results}
\end{figure}

\begin{figure}[htbp]
	\centering
	\begin{subfigure}[b]{0.48\textwidth}
		\centering
		\includegraphics[width=\textwidth]{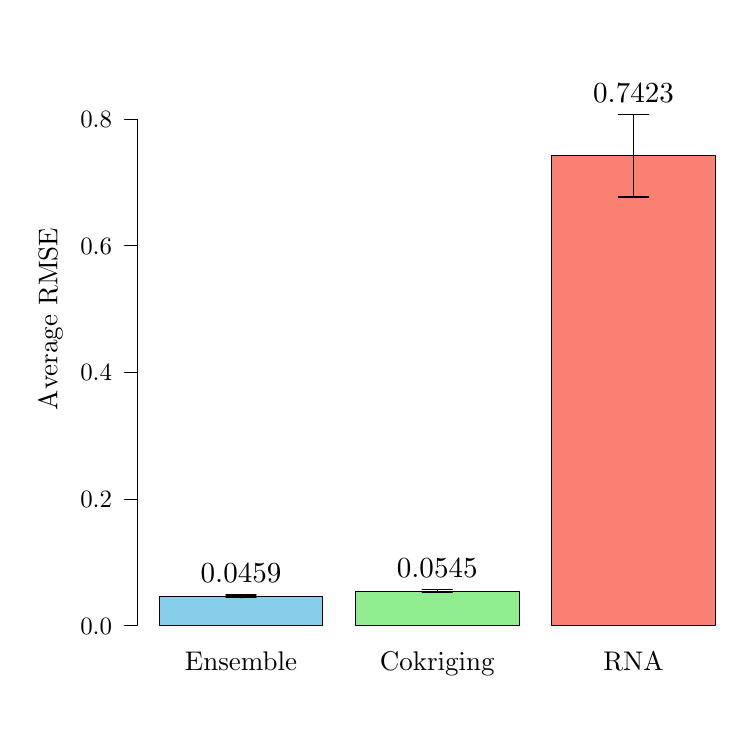}
		\caption{High sparsity, Nested}
		\label{fig:p1_ve_nest}
	\end{subfigure}
	\hfill
	\begin{subfigure}[b]{0.48\textwidth}
		\centering
		\includegraphics[width=\textwidth]{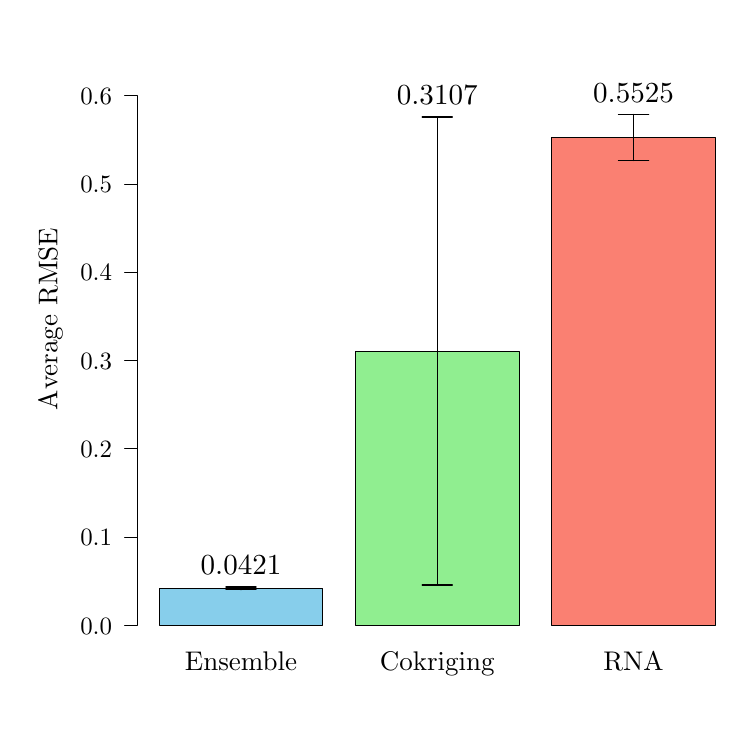}
		\caption{High sparsity, Non-nested}
		\label{fig:p1_ve_non}
	\end{subfigure}
	
	\vspace{1em}

	\begin{subfigure}[b]{0.48\textwidth}
		\centering
		\includegraphics[width=\textwidth]{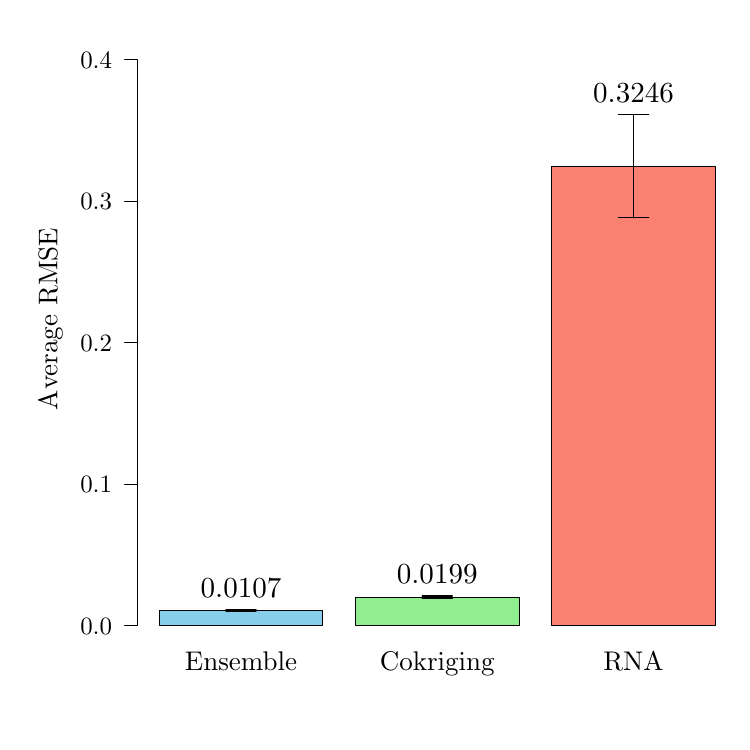}
		\caption{Low sparsity, Nested}
		\label{fig:p1_e_nest}
	\end{subfigure}
	\hfill
	\begin{subfigure}[b]{0.48\textwidth}
		\centering
		\includegraphics[width=\textwidth]{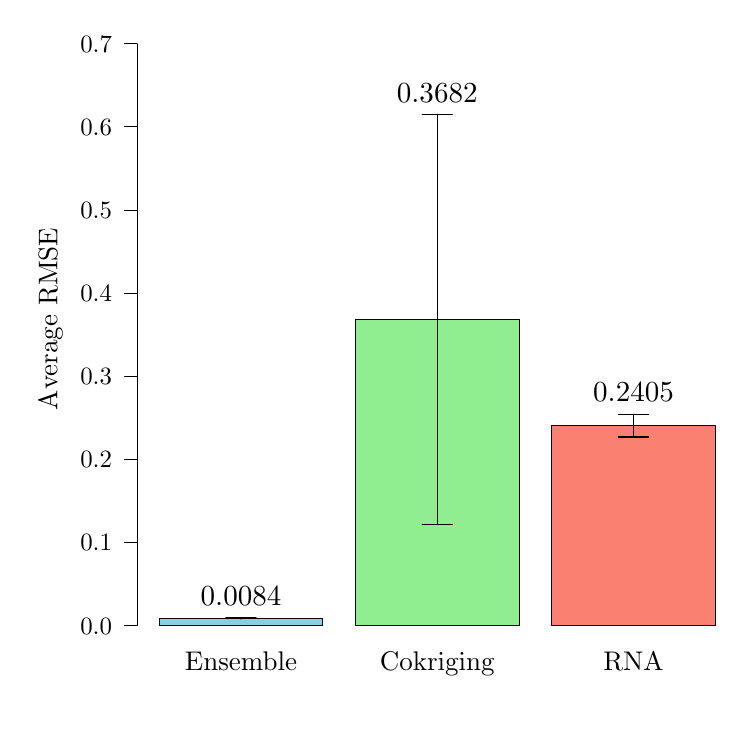}
		\caption{Low sparsity, Non-nested}
		\label{fig:p1_e_non}
	\end{subfigure}
	
	\caption{Average RMSE for the 4D Park 1 function computed over 30 different initial LHS designs. The proposed framework outperforms the competing methodologies.}
	\label{fig:park1_results}
\end{figure}

\begin{figure}[htbp]
	\centering
	\begin{subfigure}[b]{0.48\textwidth}
		\centering
		\includegraphics[width=\textwidth]{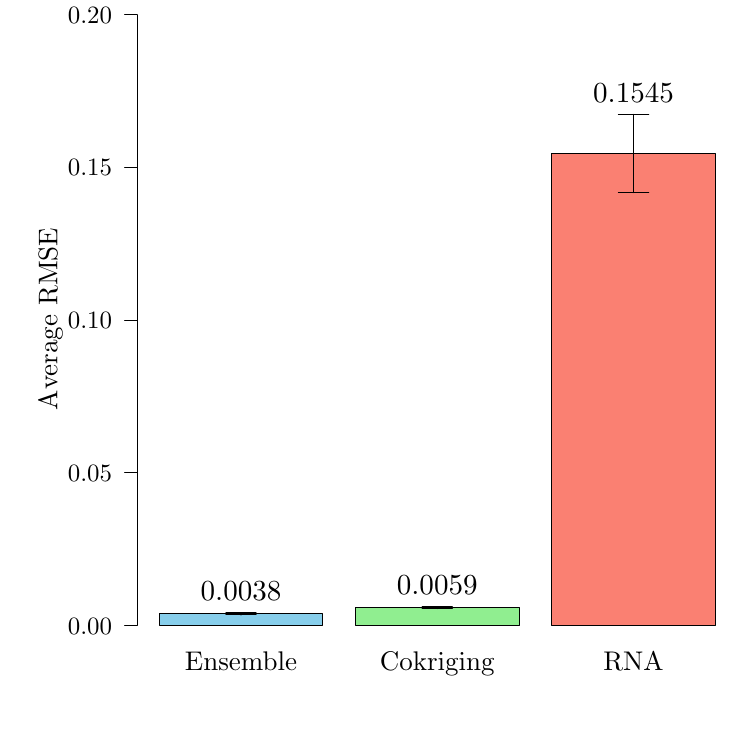}
		\caption{High sparsity, Nested}
		\label{fig:p2_ve_nest}
	\end{subfigure}
	\hfill
	\begin{subfigure}[b]{0.48\textwidth}
		\centering
		\includegraphics[width=\textwidth]{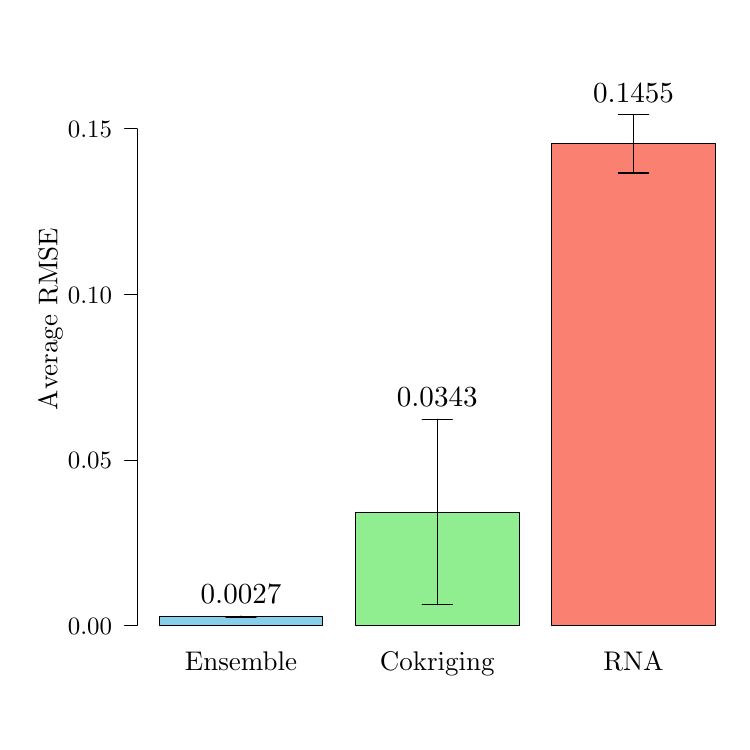}
		\caption{High sparsity, Non-nested}
		\label{fig:p2_ve_non}
	\end{subfigure}
	
	\vspace{1em}

	\begin{subfigure}[b]{0.48\textwidth}
		\centering
		\includegraphics[width=\textwidth]{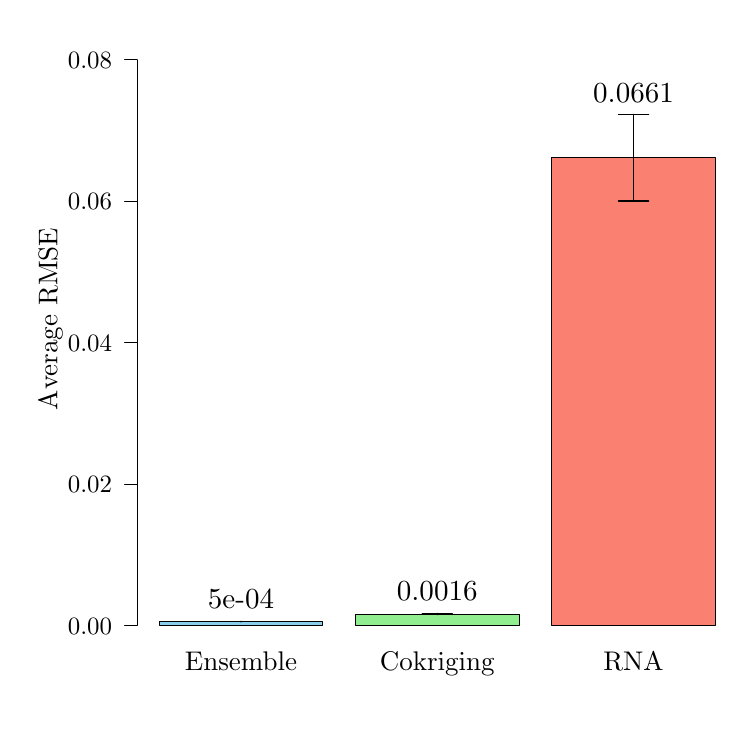}
		\caption{Low sparsity, Nested}
		\label{fig:p2_e_nest}
	\end{subfigure}
	\hfill
	\begin{subfigure}[b]{0.48\textwidth}
		\centering
		\includegraphics[width=\textwidth]{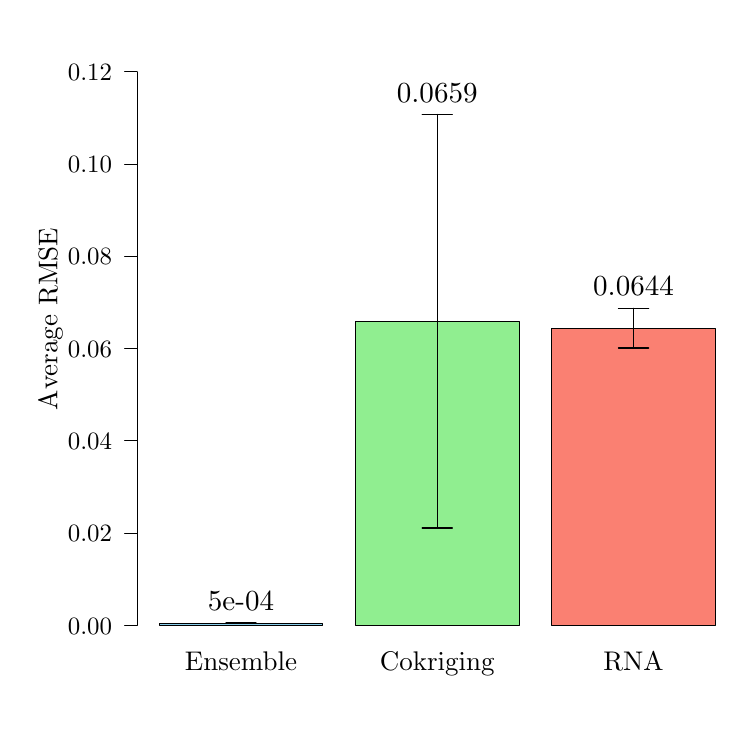}
		\caption{Low sparsity, Non-nested}
		\label{fig:p2_e_non}
	\end{subfigure}
	\caption{Average RMSE for the 4D Park 2 function computed over 30 different initial LHS designs. The results highlight the superior precision and numerical stability achieved by the proposed framework.}
	\label{fig:park2_results}
\end{figure}

\begin{figure}[htbp]
	\centering
	\begin{subfigure}[b]{0.48\textwidth}
		\centering
		\includegraphics[width=\textwidth]{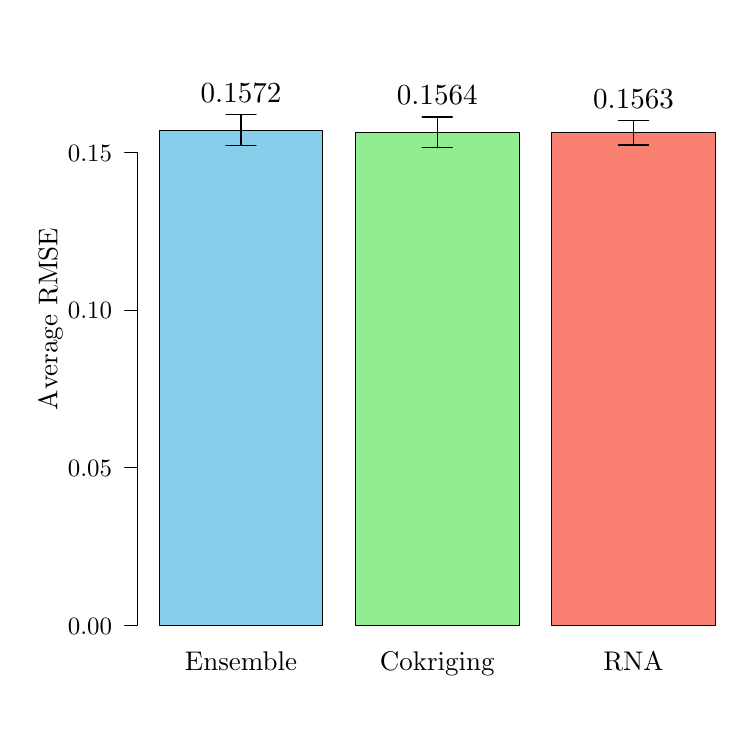}
		\caption{High sparsity, Nested}
		\label{fig:hart6_ve_nest}
	\end{subfigure}
	\hfill
	\begin{subfigure}[b]{0.48\textwidth}
		\centering
		\includegraphics[width=\textwidth]{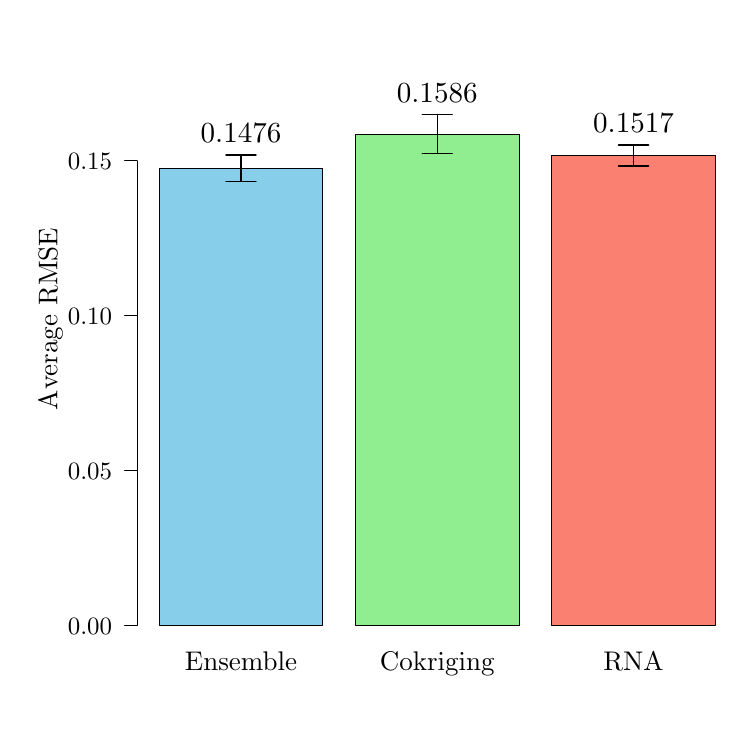}
		\caption{High sparsity, Non-nested}
		\label{fig:hart6_ve_non}
	\end{subfigure}
	
	\vspace{1em}

	\begin{subfigure}[b]{0.48\textwidth}
		\centering
		\includegraphics[width=\textwidth]{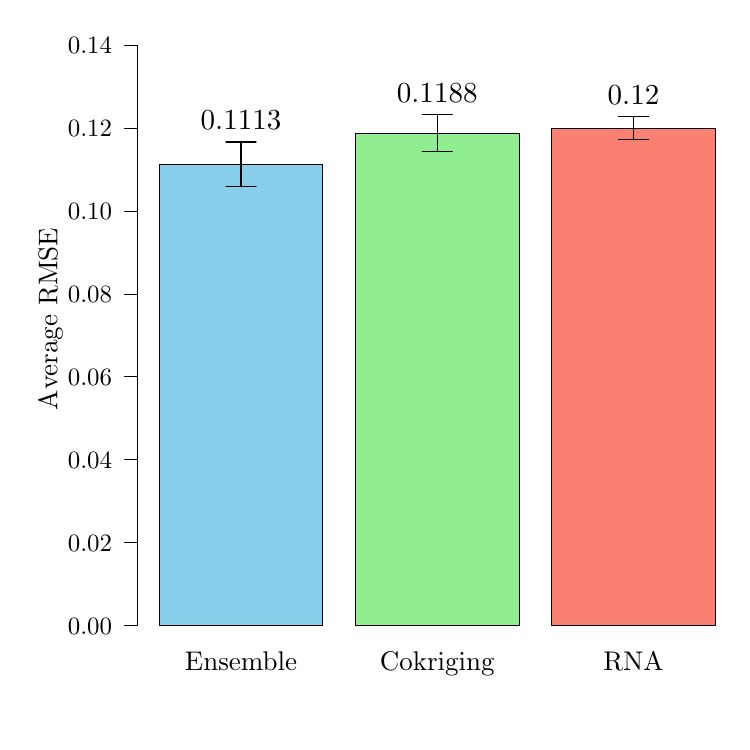}
		\caption{Low sparsity, Nested}
		\label{fig:hart6_e_nest}
	\end{subfigure}
	\hfill
	\begin{subfigure}[b]{0.48\textwidth}
		\centering
		\includegraphics[width=\textwidth]{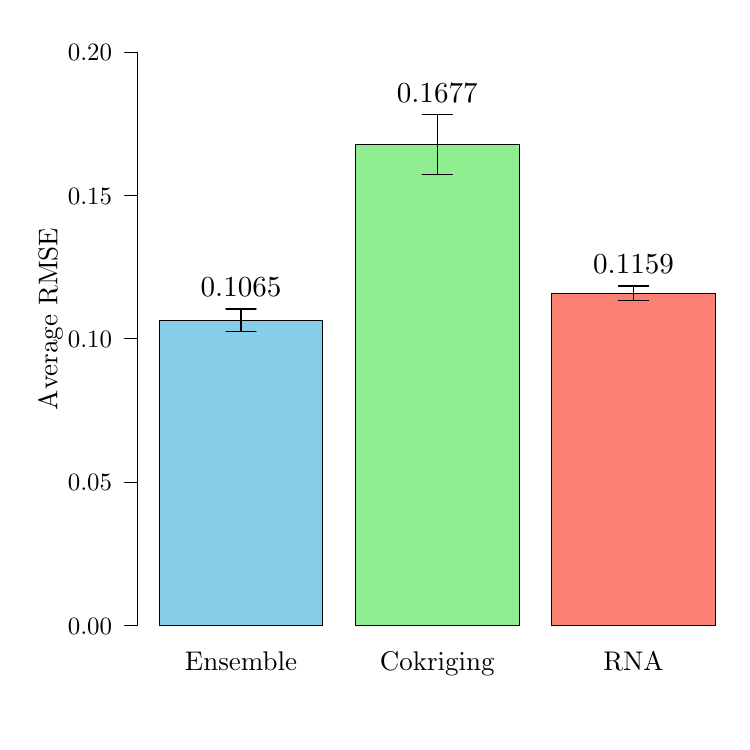}
		\caption{Low sparsity, Non-nested}
		\label{fig:hart6_e_non}
	\end{subfigure}
	\caption{Average RMSE for the 6D Hartmann function computed over 30 different initial LHS designs. While performance is comparable across all models, our approach outperforms in the low sparsity, non-nested configuration.}
	\label{fig:hart6_results}
\end{figure}

For each benchmark problem, both nested and non-nested sampling configurations are examined. In the nested case, the HF sample set is a subset of the LF set. In both configurations, model performance is assessed under two computational budgets:
\begin{itemize}
	\item High sparsity: The HF and LF sample sizes are set to $5d$ and $25d$, respectively, where $d$ is the input dimensionality.
	\item Low sparsity: The HF and LF sample sizes are increased to $10d$ and $50d$, respectively.
\end{itemize}
The samples for all cases are generated based on a maximin LHS design using the \texttt{DiceDesign} package \cite{dupuy2015}. To account for the stochastic nature of the space-filling designs, each experiment is repeated over 30 independent LHS designs. Figures \ref{fig:currin_results}-\ref{fig:hart6_results} report the average RMSE results associated with the ``Ensemble" (the proposed approach), ``Cokriging", and ``RNA" methods.

Overall, our multi-fidelity emulator shows superior performance regarding both precision and numerical stability. Its prediction accuracy is particularly high in non-nested experimental settings which are common when dealing with real-world datasets where observations at different fidelity levels are often collected independently. The framework significantly outperforms the competing approaches in approximating the Park 1 and 2 benchmark functions. In the latter, it achieves very low error levels (on the order of $5\times10^{-4}$) with a small standard error, indicating robust performance over repeated experimental designs. By employing multiple covariance kernels, the ensemble gains greater flexibility in capturing the regularity of these functions compared with the fixed Matern 5/2 kernel used in the RNA and cokriging methods.

Recursive cokriging performs slightly better than our method only on the Currin function. This behaviour can be attributed to the strong correlation between fidelity levels and the smoothness of the response, which favour autoregressive multi-fidelity models. However, the accuracy and robustness of cokriging deteriorate as the input dimensionality increases. This degradation is particularly evident in non-nested configurations (see e.g., the Park 1 and 2 functions) where results exhibit substantially higher variability. This is consistent with the robustness limitations of cokriging-based methods as discussed earlier in \Cref{sec:introduction}.

RNA generally exhibits lower predictive accuracy, except for the Hartmann function, where the average RMSE values are comparable to those of the proposed method. Nevertheless, it demonstrates greater numerical stability than recursive cokriging. Moreover, it can be observed that increasing the size of the initial training set (i.e., moving from the high to low sparsity regimes) leads to improved performance of the RNA method. Such gains are relatively modest for our approach and cokriging. However, the advantage of the proposed multi-fidelity emulator is its numerical stability, as evidenced by consistently tighter error bars across 30 independent designs.

\begin{figure}[htbp]
	\centering
	\begin{subfigure}[b]{0.48\textwidth}
		\centering
		\includegraphics[width=\textwidth]{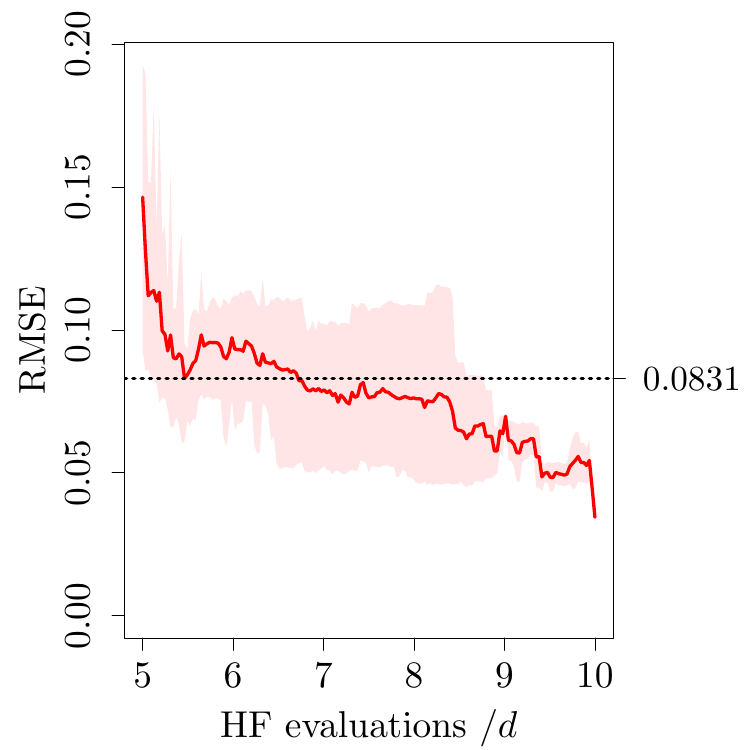}
		\caption{Currin function}
		\label{fig:currin_converg}
	\end{subfigure}
	\hfill
	\begin{subfigure}[b]{0.48\textwidth}
		\centering
		\includegraphics[width=\textwidth]{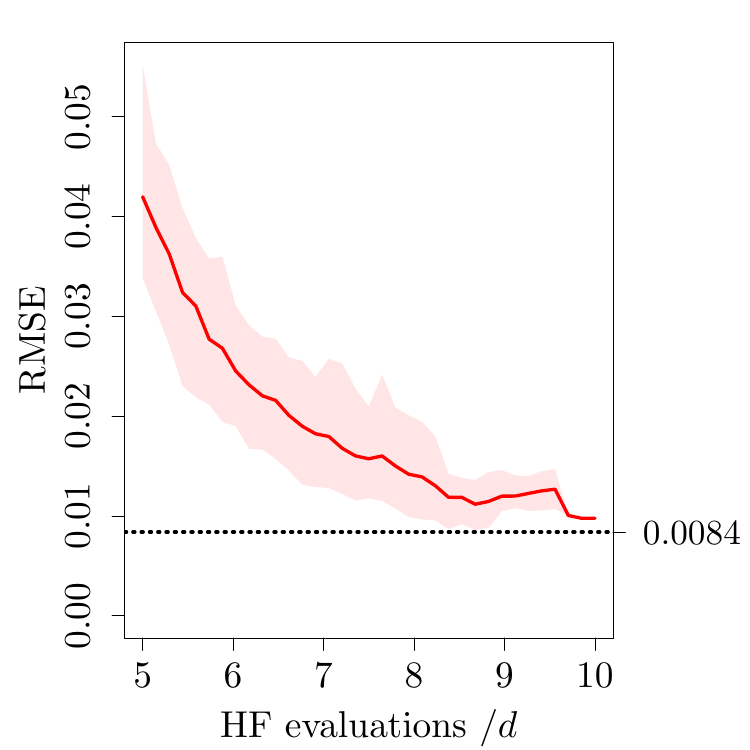}
		\caption{Park 1 function}
		\label{fig:prk1_converg}
	\end{subfigure}
	
	\vspace{1em}

	\begin{subfigure}[b]{0.48\textwidth}
		\centering
		\includegraphics[width=\textwidth]{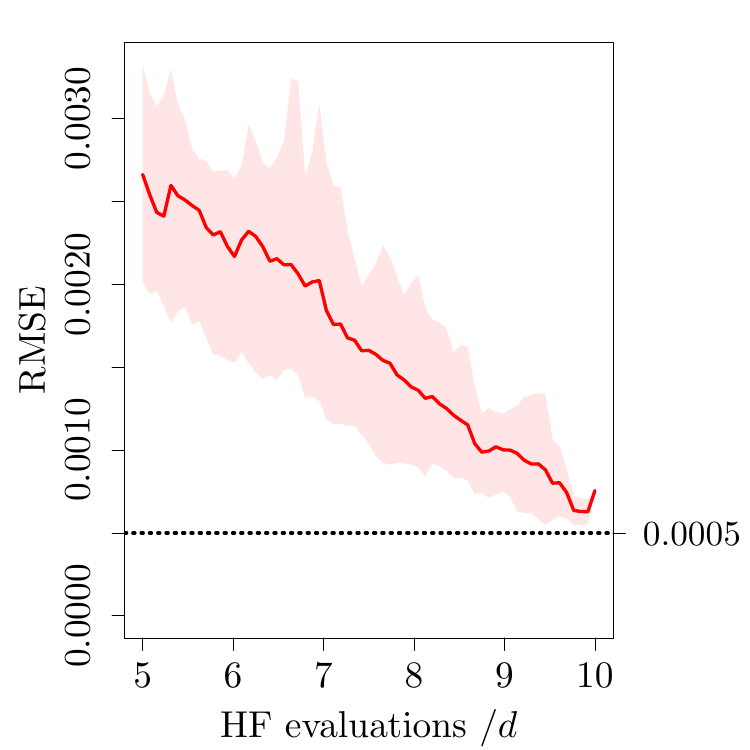}
		\caption{Park 2 function}
		\label{fig:prk2_converg}
	\end{subfigure}
	\hfill
	\begin{subfigure}[b]{0.48\textwidth}
		\centering
		\includegraphics[width=\textwidth]{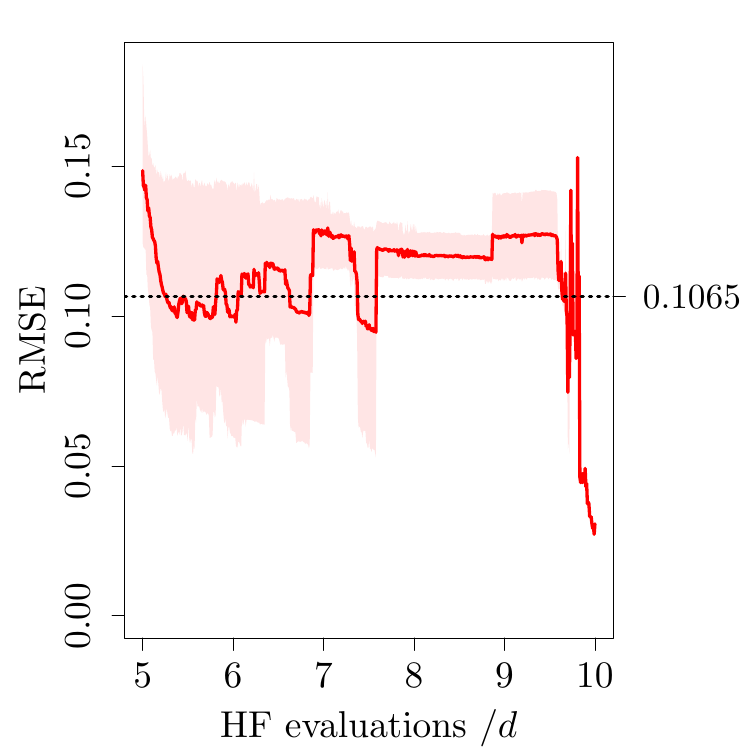}
		\caption{Hartmann function}
		\label{fig:hart6_converg}
	\end{subfigure}
	\caption{Performance of the adaptive sampling strategy using the between-model variance $\sigma^2_{\mathrm{bm}}(\bx)$ as the acquisition criterion. The red line is the average RMSE, accompanied by its corresponding uncertainty bands (shaded area). The 30 training sets from the high sparsity scenario ($5d$ HF points) are used as the initial designs and are augmented with an additional $5d$ HF samples. The horizontal dashed lines represent the average RMSE values attained by the multi-fidelity emulator under the low sparsity regime ($10d$ HF evaluations), see Figures \ref{fig:currin_results}-\ref{fig:hart6_results}.}
	\label{fig:adaptive_sampling_results}
\end{figure}

We now assess the efficiency of the adaptive design strategy, guided by the between-model variance as the acquisition criterion. The initial designs are identical to the 30 training sets used in the high sparsity scenario ($5d$). The results are presented in \Cref{fig:adaptive_sampling_results}, where the red solid line represents the average RMSE, accompanied by the corresponding uncertainty bands (shown as shaded area). The objective of this analysis is to determine whether the adaptive approach can achieve the predictive accuracy of the low sparsity regime (indicated by the dashed line) using fewer function evaluations. 

For the Currin and Park 1 and 2 cases, the RMSE decreases monotonically as the size of the training sets increases. In the Currin case, the adaptive strategy exceeds the performance of the low sparsity baseline ($0.0831$) after approximately $7d$ evaluations. For the Park 1 and 2 functions, the RMSE rapidly approaches the performance achieved in the low sparsity regime. For the Hartmann function, although the average RMSE crosses the baseline ($0.1065$) at an early stage, the shaded area indicates increases variability during the intermediate iterations of the sequential process. Toward the end of the budget ($10d$ HF evaluations), the adaptive method achieves a sharp reduction in RMSE, ultimately outperforming the low sparsity regime. These results suggest that the between‑model variance can effectively leverage model disagreement to guide the search toward informative samples that may be overlooked by space‑filling designs.

\section{Conclusions}
\label{sec:conclusion}
This paper has introduced a novel multi-fidelity emulation methodology based on ensemble learning. The ensemble consists of hierarchical kriging emulators, in which information from lower‑fidelity simulators is systematically incorporated into higher‑fidelity emulators through their trend functions. These base learners are aggregated via Bayesian model averaging, yielding a unified surrogate model with principled uncertainty quantification. This uncertainty comprises within‑model and between‑model variances, with the latter further employed in an adaptive sampling strategy to sequentially enrich the training set. The framework exhibits strong predictive performance and robustness compared with single‑model emulation approaches. Moreover, unlike conventional autoregressive and cokriging‑based methods, it does not require nested experimental designs and avoids the computational burden associated with inverting large covariance matrices. As a result, the methodology is well suited to practical settings where observations from different fidelity levels do not necessarily overlap.

Several directions for future research remain. A natural extension is multi-fidelity Bayesian optimisation \cite{kandasamy2017}, where the objective is to identify global optima rather than accurate functional approximations. Another avenue concerns the extension to problems with non-stationary or discontinuous responses by incorporating non-stationary covariance kernels \cite{mohammadi2020}. Finally, extending the proposed methodology to stochastic simulators with noisy outputs would further broaden its applicability across scientific and engineering domains.
\section*{Acknowledgements}
The author would like to thank Professor Peter Challenor for his constructive discussions.
\bibliography{biblio}
\bibliographystyle{plain}
\begin{appendices}

\section{Analytical test functions}
\label{app:test_functions}

The analytical expressions for the test functions, defined on the unit hypercube $[0, 1]^d$, are given below. The corresponding high‑ and low‑fidelity formulations are denoted by $f_{\mathrm{HF}}(\bx)$ and $f_{\mathrm{LF}}(\bx)$ respectively.

\subsection{Currin function (2D)}

\begin{align*}
	f_{\mathrm{HF}}(\bx) &= \left[ 1 - \exp\left(-\frac{1}{2x_2}\right) \right] \frac{2300x_1^3 + 1900x_1^2 + 2092x_1 + 60}{100x_1^3 + 500x_1^2 + 4x_1 + 20},\\
	f_{\mathrm{LF}}(\bx) &= \frac{1}{4} \left[ f_{\mathrm{HF}}(x_1 + \delta, x_2 + \delta) + f_{\mathrm{HF}}(x_1 + \delta, \bar{x}_2) + f_{\mathrm{HF}}(x_1 - \delta, x_2 + \delta) + f_{\mathrm{HF}}(x_1 - \delta, \bar{x}_2) \right],
\end{align*}
where $\bar{x}_2 = \max(0, x_2 - \delta)$.

\subsection{Park 1 function (4D)}

\begin{align*}
	f_{\mathrm{HF}}(\bx) &= \frac{x_1}{2} \left[ \sqrt{1 + (x_2 + x_3^2)\frac{x_4}{x_1^2}} - 1 \right] + (x_1 + 3x_4) \exp(1 + \sin(x_3)), \\
	f_{\mathrm{LF}}(\bx) &= \left( 1 + \frac{\sin(x_1)}{10} \right) f_{\mathrm{HF}}(\bx) - 2x_1 + x_2^2 + x_3^2 + 0.5.
\end{align*}

\subsection{Park 2 function (4D)}

\begin{align*}
	f_{\mathrm{HF}}(\bx) &= \frac{2}{3} \exp(x_1 + x_2) - x_4 \sin(x_3) + x_3,\\
	f_{\mathrm{LF}}(\bx) &= 1.2 f_{\mathrm{HF}}(\bx) - 1.
\end{align*}

\subsection{Hartmann function (6D)}

\begin{align*}
	f_{\mathrm{HF}}(\bx) &= -\frac{1}{1.94} \left( 2.58 + \sum_{i=1}^{4} \alpha_i \exp\left( - \sum_{j=1}^{6} A_{ij} (x_j - P_{ij})^2 \right) \right),\\
	f_{\mathrm{LF}}(\bx) &= -\frac{1}{1.94} \left( 2.58 + \sum_{i=1}^{3} \alpha_i \exp\left( - \sum_{j=1}^{6} A_{ij} (x_j - P_{ij})^2 \right) \right),
\end{align*}
where
\[
\boldsymbol{\alpha} = (1.0,\, 1.2,\, 3.0,\, 3.2),
\]

\[
A =
\begin{bmatrix}
	10 & 3 & 17 & 3.5 & 1.7 & 8 \\
	0.05 & 10 & 17 & 0.1 & 8 & 14 \\
	3 & 3.5 & 1.7 & 10 & 17 & 8 \\
	17 & 8 & 0.05 & 10 & 0.1 & 14
\end{bmatrix},
\]

\[
P = 10^{-4}
\begin{bmatrix}
	1312 & 1696 & 5569 & 124 & 8283 & 5886 \\
	2329 & 4135 & 8307 & 3736 & 1004 & 9991 \\
	2348 & 1451 & 3522 & 2883 & 3047 & 6650 \\
	4047 & 8828 & 8732 & 5743 & 1091 & 381
\end{bmatrix}.
\]

\end{appendices}

\end{document}